\RequirePackage{ifpdf}    

\documentclass[preprint,12pt,showpacs,aps,amsmath,amssymb, 
nofootinbib,superscriptaddress,hyperref,showkeys]{revtex4} 

\usepackage[utf8]{inputenc}

\usepackage{epsfig} 
\usepackage{amsmath} 
\usepackage{amssymb} 
\usepackage{wasysym} 
\usepackage{mathrsfs} 
\usepackage{graphicx} 
\usepackage{amsfonts} 
\usepackage{amsbsy} 
\usepackage{amscd}
\usepackage[normalem]{ulem} 
\usepackage{pstricks} 
\usepackage{multirow}
\usepackage{slashed} 
\usepackage{tikz}
\usepackage{hyperref}
\usetikzlibrary{decorations.pathmorphing,decorations.markings}
\newcommand{\beq}{\begin{equation}}
\newcommand{\eeq}{\end{equation}}
\newcommand{\bea}{\begin{eqnarray}}
\newcommand{\eea}{\end{eqnarray}}
\newcommand{\beas}{\begin{eqnarray*}}
\newcommand{\eeas}{\end{eqnarray*}}
\newcommand{\bi}{\begin{itemize}}
\newcommand{\ei}{\end{itemize}}

\def\tev{\,{\ifmmode\mathrm {TeV}\else TeV\fi}}
\def\gev{\,{\ifmmode\mathrm {GeV}\else GeV\fi}}

\newcommand{\mptvec}{ \, \not \! \vec{P}_T}

\allowdisplaybreaks

\begin{document}

\author{Partha Konar}
\email{konar@prl.res.in}
\affiliation{Physical Research Laboratory, Ahmedabad-380009, India}

\author{Abhaya Kumar Swain}
\email{abhaya@prl.res.in}
\affiliation{Physical Research Laboratory, Ahmedabad-380009, India}

\title{Reconstructing semi-invisible events 
in resonant tau pair production from Higgs}

\preprint{\today}

\begin{abstract}
We study the possibility of utilising the constrained mass variable, $M_{2Cons}$, in reconstructing the semi-invisible events originated from a resonant production at the LHC. While this proposal is effective for any  similar antler type production mechanism, here we demonstrate with a potentially interesting scenario, when the Higgs boson decays into a pair of the third generation $\tau$ leptons. 
Buoyed with a relatively large Yukawa coupling, the LHC has already started exploring this pair production to investigate the properties of Higgs in the leptonic sector.
Dominant signatures through hadronic decay of tau, associated with invisible neutrinos compound the difficulty in the reconstruction of such events.
Exploiting the already existing Higgs mass bound,
this new method provides a unique event reconstruction, together with a significant enhancement in terms of efficiency over the existing methods.
\end{abstract}

\maketitle


\section{Introduction}
\label{sec:intro}
The Large Hadron Collider (LHC), still lacking with its objective in confirming any clear indication of new physics beyond the Standard Model (SM), has nevertheless successfully discovered the SM like Higgs boson at 125 GeV \cite{Chatrchyan:2012ufa, :2012gk}  and also made  tremendous progress in probing different properties of this newly discovered scalar \cite{cmstwiki,atlastwiki}. Owing to the relatively large Yukawa coupling, looking for the events where the Higgs decaying into third generation $\tau$'s are the natural first step in exploring the interactions with the leptonic modes. Full event reconstruction for such event topology is especially more important, since the fermions from third generation family held the key to the electro-weak symmetry breaking, and moreover, can shed light on different aspects of the resonant state such as, coupling structure, spin, CP properties. This, in turn, can be exploited to constrain the effects coming from any possible new physics. 

The CMS collaboration recently studied \cite{Chatrchyan:2014nva} the tau pair production from the Higgs boson, at center-of-mass energy 7 and 8 TeV, corresponding to the integrated luminosity of 4.9 and 19.7 $\text{fb}^{-1}$ respectively. To explore these $\tau$ leptons, both hadronic and leptonic decay modes are considered, resulting into six different final states from the pair. This analysis reported an excess of events over the background only hypothesis, with a local significance 3.2 standard deviation correspond to the Higgs boson mass at 125 GeV. The study of $\tau$ pair final state at the LHC is rather onerous, making the significance smaller compared to other decay modes of the Higgs boson.
Difficulty lies in reconstructing the hadronic or leptonic decay modes of the tau lepton, especially in presence of the invisible neutrinos at the final state.

There are several techniques introduced for the study of $h \rightarrow \tau \tau$ process and we give an  outline as follows.
\begin{itemize}   
	\item \textit{Collinear approximation} \cite{Ellis1988221} assumes that all the decay products from the tau lepton are collinear. As a result, each neutrino, among these decay products, takes some fraction of the tau momenta. This unknown fraction can be determined by using the measured momenta of the visible particles and missing transverse momentum.
	This approximation is effective when the Higgs is produced in association with hard jet(s), boosting the tau pair system. Thus, a significant portion of events, producing the $\tau$'s back-to-back in transverse direction, remains outside the preview of this method. Therefore, the overall statistical significance from such study gets reduced.
	\item \textit{Missing mass calculator} \cite{Elagin:2010aw} replaces the collinear approximation by constructing a probability function utilising the angular information in the event, to parameterize this under-constrained system. Two remaining unsolved degrees of freedom are thus fixed whereas, rest are solved using the four constraints with $\tau$ mass-shell relations, and the missing transverse momenta. Missing mass calculator is applicable to all events, although it is computationally expensive.\footnote{While we are finalizing this article, a method~\cite{Xia:2016jec} similar to the missing mass calculator appears. This study samples all kinematically allowed values of the magnitude of invisible momentum and the visible/invisible invariant mass using their distributions  from the Monte Carlo simulations. The mass of the heavy resonance is shown to be the position of maximum probability.}
	\item \textit{Displaced vertex method} \cite{Gripaios:2012th} considers the events in which at least one of the $\tau$'s undergoes a three-prong decay. This method reconstructs the $\tau$ momenta using the secondary vertex information, together with the mass-shell and missing transverse momentum constraints.  
	This method can utilise only a small fraction of events associated with 3-prong decay of tau.
	\item \textit{Constrained $\hat{s}$ method} \cite{Swain:2014dha,Swain:2015qba} assumes the knowledge of parent mass ($m_\tau$ in present process)  and minimize the partonic mandelstam variable with respect to the unknown invisible particle momenta, taking care of missing transverse momenta constraints, to construct $\hat{s}_{min}^{cons}$ and $\hat{s}_{max}^{cons}$ variables. The new variable $\hat{s}_{min}^{cons}$ ($\hat{s}_{max}^{cons}$) exibits a sharp endpoint (threshold) exactly at the Higgs mass.
	\item \textit{Stochastic mass-reconstruction} \cite{Maruyama:2015fis} is another prescription proposed lately for the measurement of the mass of a heavy resonance decaying into tau pair. This method estimates the momenta of parent particle ($\tau$) by multiplying  the final state daughter multiplicity with the average momenta of visible daughters.
\end{itemize}

In this present work, we start with investigating yet another method based on the MAOS technique. We reconstruct the invisible momenta, followed by calculating the $\tau$ pair invariant mass. 
One expects a correct reconstruction of  heavy resonant mass if true invisible momenta was already available. In the absence of that information, the efficiency of any such reconstruction technique, in calculating the event momenta, is best represented by demonstrating the derived invariant mass.  The benefit of this MAOS method is in its applicability for all events and in a simple $M_{T2}$ based calculation for this topology with two semi-invisible tau decay chains. More importantly, it can stimulate one to use the (1+3) dimensional sister $M_{2Cons}$, which preserves all the properties of $M_{T2}$.  In addition, this new variable have the ability to utilise the on-shell mass information including that of  the Higgs and thus improve the reconstructed momentum and mass for this semi-invisible system.
Already measured Higgs mass information at the LHC is utilised in construction of the proposed variable $M_{2Cons}$, significantly improving the event reconstruction capability over the existing methods.
Although, in the present study we focus on the reconstruction of the SM Higgs boson decaying into the tau lepton pair events, this technique are in general applicable for the reconstruction of any heavy resonance producing pair of unstable particles, which subsequently decays semi-invisibly. This is typically antler \cite{Han:2009ss} type production mechanism which can be mediated either by a light or heavy scalar, or  heavy  $Z^{\prime}$ like vector boson, or some spin 2 resonance. Once the mass of the heavy resonance is finally known, the $M_{2Cons}$ can be used for a better event reconstruction and thus looking into different properties of this heavy particle.

The rest of our presentation is organized as follows. In section~\ref{sec:calc}, we make a short outline on the collinear approximation describing the principle to calculate the invisible momenta before moving into
our scenario. We introduce the $M_{T2}$ assisted method, MAOS, and once again reconstruct the events using this technique. We compare the reconstruction efficiency in both cases by constructing the $\tau$ pair invariant mass.
Knowing the mass of Higgs boson already, we thereafter introduce the (1+3) dimensional generalization $M_{2Cons}$ which, by exploiting this constraint, is expected to give an improved measurement over $M_{T2}$. Event reconstruction efficiency for longitudinal and transverse momentum components are discussed in section~\ref{sec:result} and comparison is made between these methods. We summarize and conclude in section~\ref{sec:conclusion}.

\section{Event reconstruction methods}
\label{sec:calc}

The collinear approximation is one of the most popular method used for the reconstruction of invariant mass $m_{\tau\tau}$ in semi-invisible decay of  $h\rightarrow \tau\tau$ process. The primary assumptions associated with this method is, all decay products of $\tau$ lepton are collinear and the source of missing transverse momenta is due to the neutrinos only.
Following the above mentioned presuppositions, the visible decay products from each $\tau$ take some fraction of the respective $\tau$ momenta, $f_i$ with $i = 1,2$. So in a particular event these two unknown fractions can be solved using missing transverse momenta constraints. As a result, full reconstruction of the event is possible.  But when the Higgs boson is produced with small (zero) transverse momenta the two $\tau$ leptons are going back-to-back in transverse direction, making the reconstruction of  $\tau$ momenta impossible. The situation can be surpassed  if the Higgs boson is  produced with sufficient non-zero transverse momenta, that may come from associated production of initial state radiation (ISR) or extra hard jet(s).

\begin{figure}[t]
	\centering
	\includegraphics[scale=0.32,keepaspectratio=true]{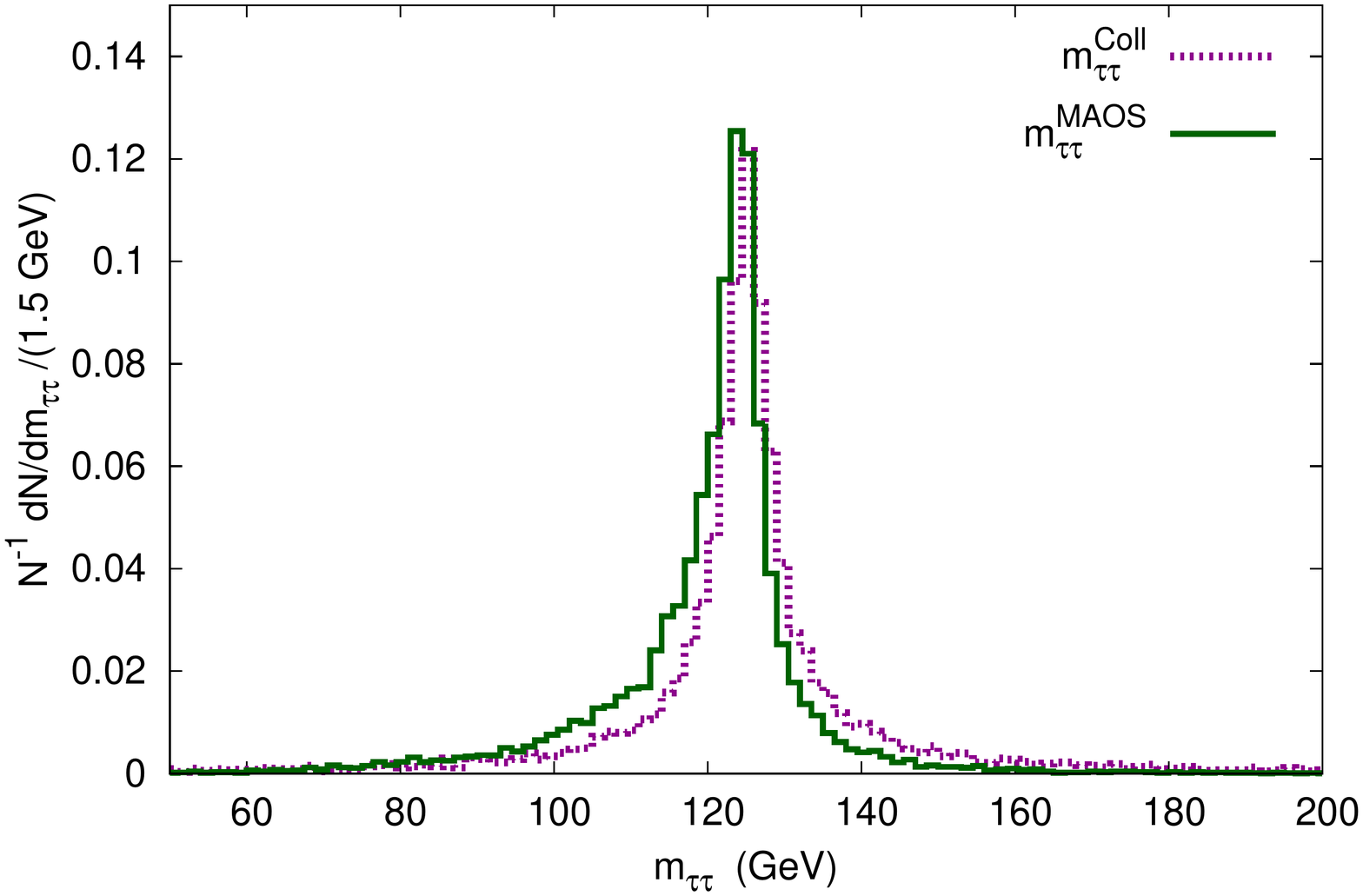}
	\caption{The purple-dotted histogram delineates the normalized distribution of $\tau$ pair invariant mass $m_{\tau\tau}$, calculated using the collinear approximation. Similarly, in the same plot, the green-solid histogram describe the same quantity utilising the MAOS momentum reconstruction method. The peak position of both these  distributions are at the Higgs mass providing with comparable efficiency.}
	\label{fig:mtautaucoll}
\end{figure}

The neutrino momenta in terms of visible particle momenta  is $\vec{q}_i = \vec{p}_{\tau_i} - \vec{p}_{i} = F_i \vec{p}_{i}$, where $F_i = \frac{1}{f_i} - 1$. 
$p_{\tau_i}$ are momenta of $\tau$'s in $h\rightarrow \tau\tau$ process, while $p_{i}$ and  $q_i$ are the final visible momenta and neutrino momenta respectively from each of these $\tau_i$ decay. We are following this momentum convention  throughout this draft.  The two unknown fractions $f_i$ can be solved using the following  two transverse equations
\begin{eqnarray}\label{missingpt}
	{\slashed{\vec{p}}_T} = \sum_i F_i \vec{p}_{iT}.
\end{eqnarray}
The solutions for $f_i$ are,

\begin{eqnarray}
	f_1 = \frac{1}{1 + r_2}\label{x1coll}; \,\,  f_2 = \frac{1}{1 + r_1}\label{x2coll}
\end{eqnarray}
with $r_i = |{\frac{\slashed{p}_y p_i^x - \slashed{p}_x p_i^y}{p_1^y p_2^x - p_1^x p_2^y}}|$ are positive dimensionless ratios constructed in terms of measured momentum combination.
The invariant mass of $\tau\tau$ system using the collinear approximation is $\frac{m_{vis}}{\sqrt{f_1f_2}}$, where $m_{vis}$ is the total invariant mass of all the visible particles.
In figure~\ref{fig:mtautaucoll}, we have presented the normalized distribution for the invariant mass $m_{\tau\tau}$ (in purple-dotted histogram), calculated using the collinear approximation. The peak of the distribution is exactly at the Higgs mass.

Now it is evident from the eqn.~\ref{x1coll} that when the two $\tau$'s are back-to-back in transverse direction the collinear approximation fails to work. Similar argument can be realized in terms of azimuthal angle \cite{Elagin:2010aw}. 
In this present analysis, parton level simulated events\footnote{
	Note that the default setting in PYTHIA 8 generates both the hadronic and leptonic decay of tau preserving spin correlations based on fully modeled $\tau$ lepton decay.  Along with that, in our present analysis, we generate parton-level simulated events keeping the hadronization option off, leaving the realistic analysis including the particle identification and detector level simulation for future work.} for  $h\rightarrow \tau\tau$  are generated  along with the ISR jet(s) using PYTHIA 8~\cite{Sjostrand:2007gs} and there by making a suitable momentum configuration for the collinear approximation to work. 
%
One can also notice from the histogram that the collinear approximation shifts the reconstructed invariant mass towards a higher value, and also develops a tail at larger invariant mass. This is consequence of some of the events coming with soft (ISR) jets. This tail becomes rather significant once realistic events with measurement errors are also included \cite{Elagin:2010aw}.
Subsequently, the information of the heavy resonance mass is utilized in addition to the collinear approximation for the full reconstruction of the tau pair events~\cite{Anderson:1992jz}. This additional constraint improves the reconstruction of tau lepton momenta. This technique is effective even if the tau leptons are produced nearly back-to-back in transverse plane as happens for a significant portion of events. 
In the present analysis we have already considered the Higgs boson produced with sufficient transverse momenta balanced by ISR jet(s). 
So this tail feature in the distribution (as  noted in figure~\ref{fig:mtautaucoll}) is not prominent and this provides an estimate of efficiency which one expects after using the resonant mass constraint.

We now move to examine the ability of MAOS \cite{Cho:2008tj, Park:2011uz}, a $M_{T2}$-based ($M_{T2}$ assisted on-shell) method, for the full reconstruction of the tau pair events. The $M_{T2}$ \cite{Lester:1999tx, Cho:2007qv} is defined as the maximum of transverse mass, constructed for each $\tau$ using missing transverse momentum constraints, minimised over the invisible particle momenta. In the MAOS method  transverse momenta of the invisible particles are assigned to the values that gives this minimization. The longitudinal momentum is further determined using the two mass-shell conditions $(p_i + q_i)^2 = m_{\tau_i}^2$. Hence, the MAOS method reconstructs the full event with a four fold ambiguity, arises because of the quadratic mass-shell constraints.

The mass of heavy resonance can be constrained by calculating the invariant mass of both the $\tau$'s, $m_{\tau\tau}^{MAOS}$, with their assigned MAOS momenta, $p_{\tau_i}^{MAOS}$.
Where $p_{\tau_i}^{MAOS} = p_i + q_i^{MAOS}$ and all the four fold ambiguity is taken into account by the superscript $MAOS$. In the same figure~\ref{fig:mtautaucoll}, we have also shown the normalized distribution (green-solid curve) considering this MAOS reconstruction and recognize that both the methods display equal level of efficiency in reconstructing the invariant mass.
Note that 
we utilize the same Higgs data associated with additional jet\footnote{Therefore we actually used the subsystem \cite{Burns:2008va} based $M_{T2}^{sub}$ but to avoid notational cumbersome we simply write it $M_{T2}$ (also similarly in case of $M_2$) in this whole article.} for this analysis. 
That was essential for the collinear approximation to work, but 
MAOS method can be applied to all momentum configuration of the considered process, leading to a statistical advantage over the collinear approximation. Same argument is also true for our proposed method which we would discuss next. 

We now wonder whether the event reconstruction can be improved using MAOS along with the heavy resonance mass shell constraint. Since MAOS assign the transverse momenta from the minimisation of $M_{T2}$, a (1+2) dimensional variable, $q_T^{MAOS}$ can not be constrained by the heavy resonance mass. But MAOS uses (1+3) dimensional mass shell constraints to assign longitudinal momentum. So the heavy resonance mass shell constraint may be used along with parent mass to get longitudinal momentum. But the full event reconstruction may not be improved.
We now look at the possibility to construct the mass variable where this mass constraint can be used  more inclusively.

\begin{figure}[t]
	\centering
	\includegraphics[scale=0.32,keepaspectratio=true]{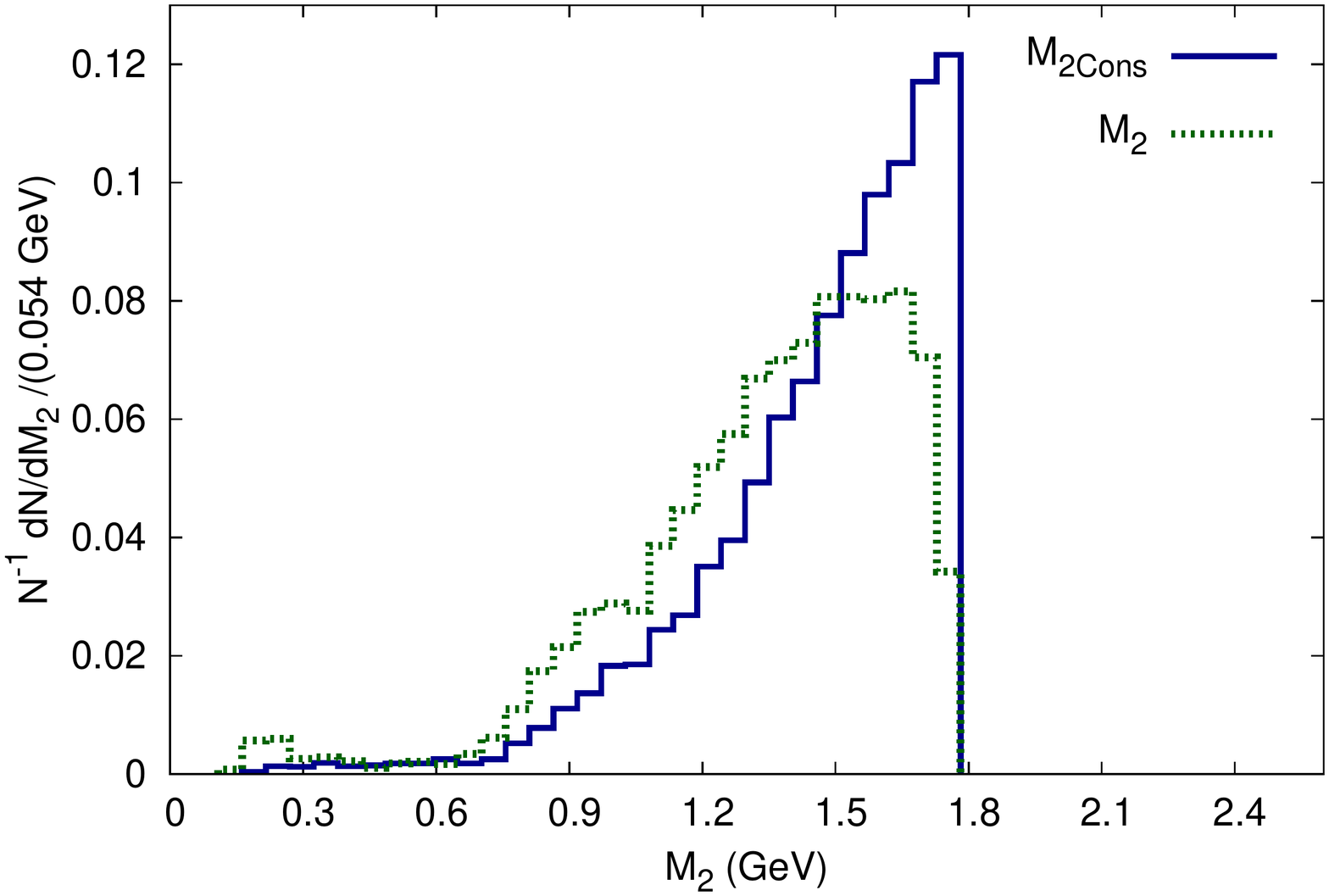}
	\caption{Normalized distributions for (1+3)-dimensional mass constraining variables in the process when Higgs decays semi-invisibly through $\tau$ pair production. 
		Green-dotted histogram describes the $M_2$ distribution considering hadronic decays of $\tau$, consists of both 1-prong and 3-prong decays. As expected, the figure shows that the endpoint of this distribution is at $m_{\tau}$ mass. Similarly, the blue-solid histogram representing the $M_{2Cons}$ and also has endpoint at the same point. However, endpoint in this case is populated with much larger number of events. 
	}
	\label{fig:m2m2cons}
\end{figure}

We shift our focus from transverse mass variables to $M_2$, which is a (1+3)-dimensional variable \cite{Barr:2011xt, Barr:2012im, Cho:2014naa} used for the determination of mass of the unstable particle, produced in pair and decaying semi-invisibly. This variable can use the longitudinal momentum component information which enables it to use available mass-shell constraints of resonance particle. This capability was lacking in its predecessor $M_{T2}$, although, this is an efficient variable for mass and spin measurement. $M_2$ is defined as
\begin{equation}\label{m2}
	M_{2} \equiv  \min_{\substack{\vec{q}_{1}, \vec{q}_{2} \\  
			\left\{   \substack{   \vec{q}_{1T} + \vec{q}_{2T} = {\mptvec} }   \right\}  }}     \left[  \max_{i =1, 2}   \{M^{(i)}(p_{i}, q_{i}, m_{v_i}; \tilde{m}_{\nu})\} \right]
\end{equation}
with
\begin{equation}\label{mass}
	M^{(i)} = m_{v_i}^2 + \tilde{m}_{\nu}^2 + 2(E^{vis(i)}E^{inv(i)} - \vec{p}_{i}.\vec{q}_{i}).
\end{equation}
After executing the additional minimisation over the  z-components of invisible particle momenta, $M_2$ comes out to be exactly equal to its (1+2)-dimensional analog, $M_{T2}$ \cite{Cho:2014naa}. Hence, all the properties of $M_{T2}$ transmits to  its successor $M_2$ with additional advantages, accommodating  the on-shell mass constraints 
as discussed earlier. One important property of the $M_2$ (or $M_{T2}$) is that, by construction, this quantity needs to be less than or equal to the unstable parent mass, $m_\tau$, given a massless  invisible daughter hypothesis ($\tilde{m}_{\nu} = 0$). So, over many events, the distribution of $M_2$ has an endpoint exactly at the true mass of mother particle.
The distribution of $M_2$ mass variable considering the semi-invisible decay of tau pair is displayed in figure~\ref{fig:m2m2cons} in green-dotted histogram. It is clear from the figure that the endpoint of $M_2$ is at $m_{\tau}$ as expected. We have considered only the hadronic decays of $\tau$'s  encompassing both the 1-prong and 3-prong decays. The tau lepton has a branching ratio of around $66\%$ for hadronic decays of which 1-prong and 3-prong decay accounts for $50\%$ and $15\%$ respectively, while rest are other hadronic decays. 
Although, the leptonic decay modes towards electron and muon production can have a considerable branching ratio together with a relatively better energy resolution, we have not considered these decay modes in the present analysis. With associated three to four neutrinos in the final states event reconstruction is impossible unless one invoke some kind of approximation.

Equipped with the Higgs mass ($m_h$), already measured  in the first run of the LHC, $M_2$ can be further improved with this constraint and proved to be useful in providing the invisible particle momenta in great efficiency.
We define a variable $M_{2Cons}$ which was first proposed in the ref. \cite{Konar:2015hea}%
\footnote{$M_{2Cons}$ was proposed for a general antler type topology for determination of the parent and invisible daughter mass. This constrained variable assumes the knowledge of heavy resonance mass. Importantly,  $M_{2Cons}$ display a new kink behavior exactly at the true mass of the parent and daughter in the correlation curve. So by identifying the position of kink one can determine both the parent and daughter mass simultaneously. Another key feature of this variable is that it gives a well correlated and unique momentum reconstruction for the invisible daughter. In the present study since we know the mass of tau lepton and that of neutrino, $M_{2Cons}$ is used for the event reconstruction purpose.}. 
Mathematically, it is defined as,
\begin{equation}\label{m2Cons}
	M_{2Cons}
	\equiv  \min_{\substack{\vec{q}_{1}, \vec{q}_{2} \\   
			\left\{   \substack{  \vec{q}_{1T} + \vec{q}_{2T} = {\mptvec} \\  (p_1 + p_2 + q_1+ q_2)^2 = m_h^2 }  \right\}   }} 
	\left[ \max_{i =1, 2}  \{ M^{(i)}(p_{i}, q_{i}, m_{v_i}; \tilde{m}_{\nu}) \} \right],
\end{equation}
with $M^{(i)}$ is expressed as in eqn.~\ref{mass}.

The Higgs mass-shell condition further constrains the invisible momenta and thus making the allowed phase space shrink to a comparatively smaller region.\footnote{The detailed discussion on the squeezed phase space under the influence of additional mass-shell constraint can be followed from the ref.~\cite{Konar:2015hea}.} Hence, the derived value of $M_{2Cons}$ comes out to be greater than or equal to $M_2$. Both of these quantities are  bounded by the tau mass, satisfying the relation $M_2 \le M_{2Cons} \le m_{\tau}$ considering each event. Consequently, more number of events move towards the endpoint in the distribution of $M_{2Cons}$ in comparison to $M_2$. The distribution of $M_{2Cons}$ is also shown in the figure~\ref{fig:m2m2cons} in blue-solid curve which clearly demonstrates that the constrained variable $M_{2Cons}$  exhibits a very sharp endpoint with a large number of events present there, enabling a better mass measurement and momentum reconstruction which we discuss now.

\section{Correlation of reconstructed momenta with true neutrino momenta}
\label{sec:result}

\begin{figure*}[!t]     
	\includegraphics[scale=0.27,keepaspectratio=true]{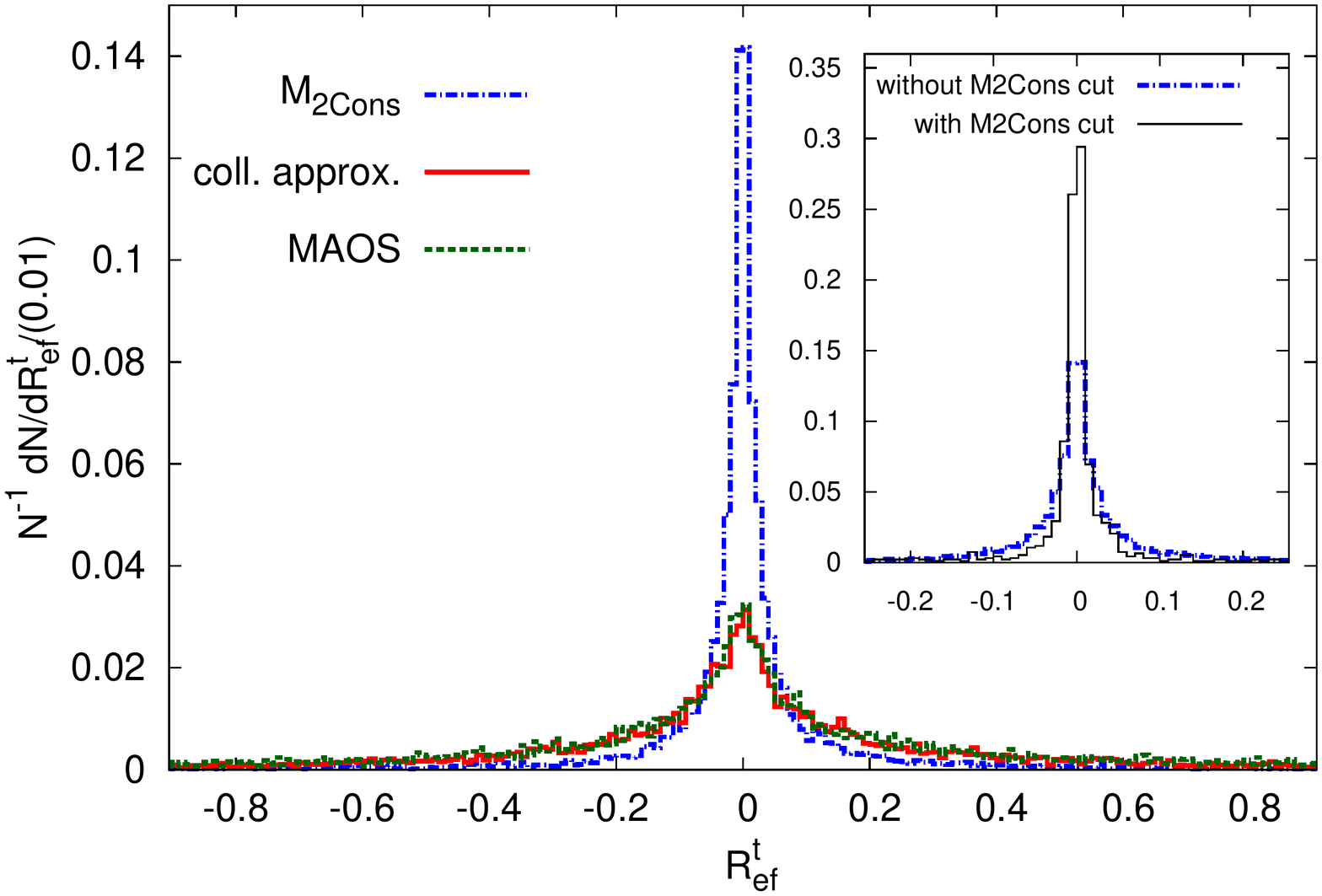}
	\includegraphics[scale=0.27,keepaspectratio=true]{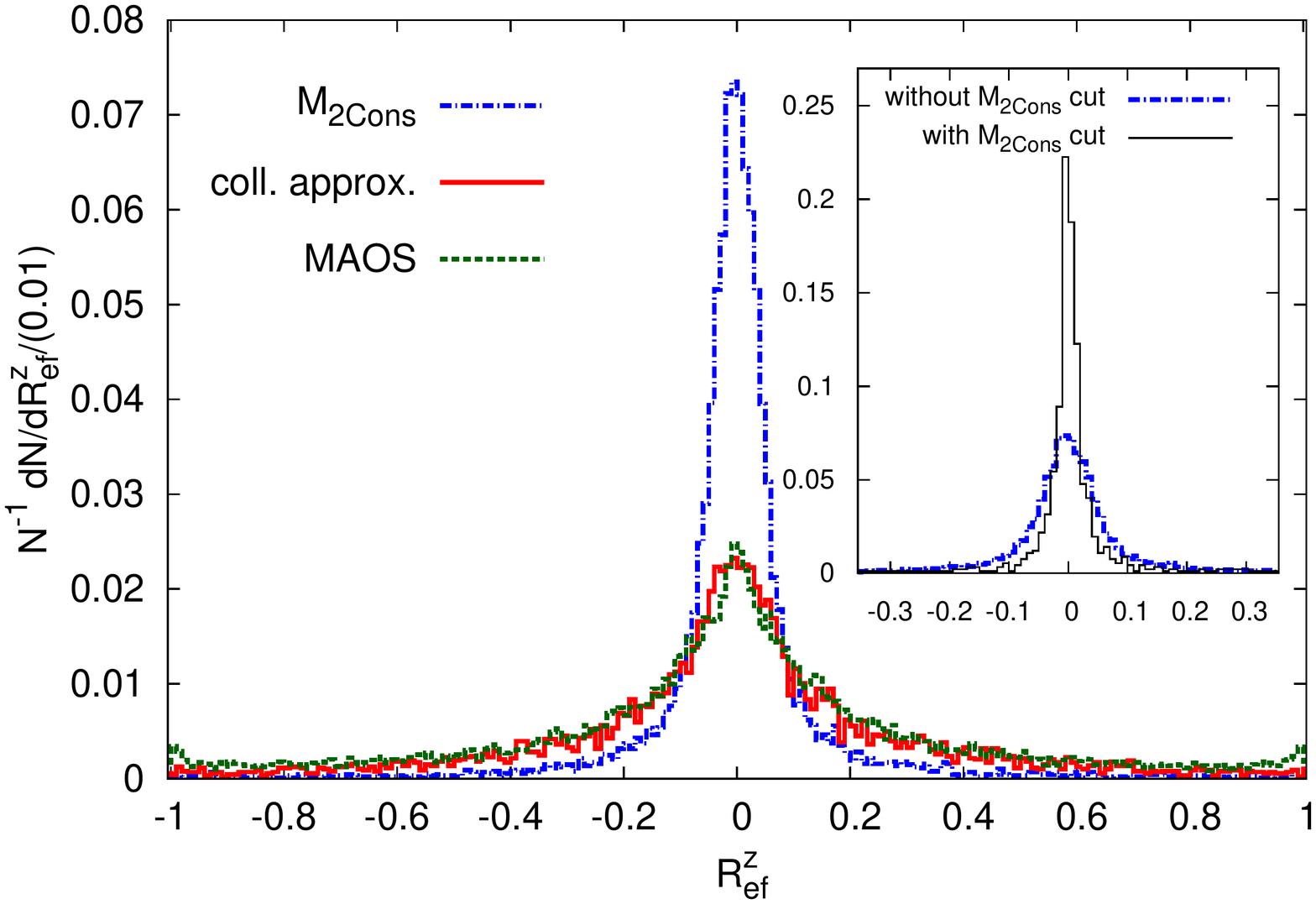}
	\caption{Efficiency of different methods for reconstructing the events coming from semi-invisible Higgs boson decay through $\tau$ pair production, after considering hadronic decays of $\tau$, consists of both 1-prong and 3-prong decays.
		Deviation of the reconstructed momenta from the true invisible momenta are parameterized using two variables (Left plot) $R_{ef}^t$ for transverse part and (Right plot) $R_{ef}^z$ for longitudinal momentum. The distributions of these variables utilising  the $M_{2Cons}$, collinear approximation and MAOS method are exhibited in blue-dashed-dotted, red-solid and greed-dotted lines respectively. The event reconstruction capability of the collinear approximation and MAOS method are of same order (as also detected in figure~\ref{fig:mtautaucoll}) while $M_{2Cons}$, with the help of additional mass constraint, is showing significant improvement. (Inset plots) Efficiency of reconstruction for both the transverse part and longitudinal momentum comparing with and without using $M_{2Cons}$ cut. The $10\%$ of events are selected towards the upper endpoint of $M_{2Cons}$ and the  reconstructed momenta with these events are found to be highly correlated with the true momenta of the invisible particle in comparison to the full data set.}
	\label{fig:errorinq1}
\end{figure*}

In this section, we parametrise the efficiency of event reconstruction for  $h\rightarrow \tau\tau$ event using different methods including collinear approximation and argue the effectiveness in using the $M_{2Cons}$ in calculating\footnote{We have calculated the mass variable $M_2$ and $M_{2Cons}$ using constrained optimization method in Mathematica. Towards the end of this analysis, a generic package, OPTIMASS[24], for the calculation of mass variables appeared which is a minuit2 based method. OPTIMASS can also be used for the calculation of $M_{2Cons}$ 
	with a simple modification of the constraint in the examples demonstrated there.} the invisible particle momenta. Event reconstruction of such events are of particular interest for spin, polarization, coupling measurement and CP symmetry studies~\cite{Harnik:2013aja,Berge:2011ij}. We use two dimensionless parameters, $R_{ef}^t$ for transverse ($x$ or $y$ component)  and $R_{ef}^z$ for longitudinal momenta, to determine the efficiency of  event reconstruction.
\begin{eqnarray}
	R_{ef}^t &=& \frac{\Delta q_t}{|q_t^{True}|} = \frac{q_t^{Reco} - q_t^{True}}{|q_t^{True}|}, \\
	R_{ef}^z &=& \frac{\Delta q_z}{|q_z^{True}|} = \frac{q_z^{Reco} - q_z^{True}}{|q_z^{True}|}. 
\end{eqnarray}
By construction, the variable $R_{ef}^t$ and $R_{ef}^z$ acquire zero value once the reconstructed momenta matches with the true invisible particle momenta in a particular event. Hence, the efficiency of any reconstruction method is judged depending on the number of events having vanishing values of $R_{ef}^t$ and $R_{ef}^z$. In other words, sharper the peak of the distribution coupled with higher number of events, the better is the efficiency of reconstruction.
It is straightforward  to calculate $R_{ef}^t$ and $R_{ef}^z$ for collinear approximation, once the fractions $f_i$ are known. In figure~\ref{fig:errorinq1}, the left plot showing the distribution of $R_{ef}^t$ utilising the collinear approximation, MAOS and $M_{2Cons}$ methods respectively. Similarly, the right plot displays the distribution of $R_{ef}^z$ for all these  methods. The reconstructed momenta using $M_{2Cons}$ are shown to be unique and very well correlated with the true momenta of the invisible particle. It is evident from the figure that $M_{2Cons}$ gives significant improvement in event reconstruction compared to the collinear approximation and MAOS method.

The efficiency of reconstruction for both the  transverse part and longitudinal momentum from $M_{2Cons}$ can be improved further by selecting events near the upper endpoint of $M_{2Cons}$ distribution. Although, the additional constraint in $M_{2Cons}$ already pushed the values towards the endpoint and improved over conventional MAOS calculation, one can still use a $M_{2Cons}$ cut to improve the event reconstruction with higher statistics in comparison to MAOS as evident from figure~\ref{fig:m2m2cons}. In the inset plots of figure~\ref{fig:errorinq1}, we compare the improvement in reconstruction efficiency using the $M_{2Cons}$ selection. Only 10$\%$ of events are selected towards the upper endpoint of $M_{2Cons}$ and the reconstructed momenta are proved to be highly correlated with the true momenta of the invisible particle in comparison to the full data set, at a price of event statistics.

\section{Summary and Conclusion}
\label{sec:conclusion}
With increasing Higgs data it is exciting time to verify and validate different properties of Higgs including the exploration of minuscule couplings with leptonic sector. In this article we study different methods to reconstruct the events related to the important third generation tau lepton. Along with the popular machinery using collinear approximation, we also looked into the efficiency in MAOS method. We further examined the effectiveness of another variable in $M_2$ class accommodating the Higgs mass information in the analysis.  $M_{2Cons}$ can thus shown as a capable variable to provide a very accurate and unique reconstruction of such events.

\bigskip
\acknowledgments
This work was funded by Physical Research Laboratory (PRL), Department of Space (DoS), India.


\bibliographystyle{unsrt}
\bibliography{bibliography}

\end{document}